# Spectroscopic identification of habitable extra-solar planets

Eyal Schwartz, Stephen G. Lipson and Erez N. Ribak

*Physics Department, Technion—Israel Institute of Technology, Haifa, Israel, 32000*

An Earth-like extra-solar planet emits light which is many orders of magnitude fainter than that of the parent star. We propose a method of identifying bio-signature spectral lines in light from known extra-solar planets based on Fourier spectroscopy in the infra-red, using an off-center part of a Fourier interferogram only. This results in superior sensitivity to narrower molecular-type spectral bands, which are expected in the planet spectrum but are absent in the parent star. We support this idea by numerical simulations which include photon and thermal noise, and show it to be feasible at a luminosity ratio of $10^{-6}$ for a Sun-like parent star in the infra-red. We also carried out a laboratory experiment to illustrate the method. The results suggest that this method should be applicable to real planet searches.

In this work we propose a method of improving the chances of identifying a habitable Earth-like extra-solar planet using its spectral signature in the infrared. In Fourier transform spectroscopy (Bell, 1972) the spectrum of a source is investigated using a two beam interference pattern with variable path difference. A spectral feature (line or band) with center wavenumber $k$ and width $\delta k$ gives rise to oscillations having period $2\pi/k$ which decay within a path difference of $\sim 1/\delta k$. Thus, sharper spectral lines result in interferometric details at larger path differences. Moreover, from Parseval's theorem, which states that the energy in the interferogram due to a specific spectral feature is equal to that in the feature itself, a sharper line must correspond to a proportionally wider and weaker contribution to the interferogram in a given range. As a result, if we know in advance the range of line- or band-widths of spectral features which interest us, we can choose an appropriate part of the Fourier interferogram which optimizes the detectivity of such spectral features. In this way, available observing time can be used to collect the data most likely to be significant. For example, detection of the narrower line in Figure 1 will have a better signal-to-noise ratio if we only measure the marked section (Ribak 2006).

An advantage of a Fourier spectrometer compared to a dispersion spectrometer is that for this purpose it is not necessary to have specific information on center wavelengths of the bands expected, because the molecules which have evolved will certainly be different than those on Earth. Most organic molecules have CH bonds which have absorption band widths of order 3 $cm^{-1}$, although the exact center wavelength depends on the bond strength and the molecular structure. A Fourier spectrometer can be designed to use the off-center region of the interferogram which is most sensitive to bands of this width, so that the signal-to-noise ratio is improved significantly (Schwartz et al. 2011; Lipson et al. 2011). For example, in the Earth's atmosphere the only common molecule with CH bonds is $CH_4$, which has abundance of about $10^{-6}$. As a result, concentrating the sampling to the interferogram region where bands of width 3 $cm^{-1}$ dominate, we can perform a spectral analysis which has a significant advantage in sensitivity over straightforward Fourier spectroscopy using the complete interferogram. Since in general

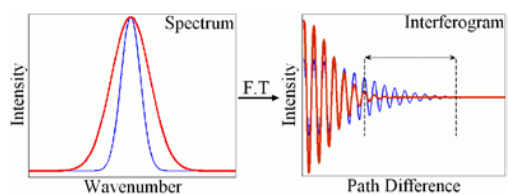

Figure 1. ((Left) Two spectral features with a common central wavenumber and different widths. (Right) The interferogram of both spectral features with the chosen part for differentiating the sharper narrow feature from the broad one.

Fourier spectroscopy and dispersion spectroscopy have the same signal-to-noise ratio when used with photon limited detectors and negligible broad-band background radiation or detector noise, this method should show an advantage over dispersion spectroscopy, unless a very limited range of wavelengths is scanned in the available observation time. If detector noise is significant, Fourier spectroscopy has the known Fellget advantage over dispersion spectroscopy (Bell 1972). We show below that use of a cold field stop and adaptive optics can remove sufficient background radiation to make the system photon limited. An extension of the general idea could be applied to a more complex shape (e.g. a multiplet) for which an optimized sample of the interferogram could be defined.

We have studied in detail the application of the above proposal to detection of bio-signatures in the spectrum of a star with an unresolvable planet, where the emission from the planet competes with the much stronger emission from the star. We consider the spectral signature of a star with a habitable extra-solar planet (exoplanet). This is expected to contain spectral details resulting from the star, from starlight reflected by the planet and from self-emission by the planet itself. If the planet has a bio-signature its reflectance and self-emission should contain molecular bands (Table 1).

| Table 1. modeled extra solar bio-signature spectral lines for the mid-IR range | | |
|---|---|---|
| Molecule | Spectral line peak $cm^{-1}$ ($\mu m$) | Spectral line width $cm^{-1}$ ($\mu m$) |
| $H_2O$ | 512 (19.51) | 2.66 (0.1) |
| $N_2O$ | 1285 (7.78) | 2.2 (0.013) |
| $CH_4$ | 1280 (7.8) | 2.86 (0.017) |
| $O_3$ | 1040 (9.6) | 3.89 (0.036) |
| Chlorophyll a - $C_{55}H_{72}MgN_4O_5$ | 1478 (6.76) | 3.5 (0.016) |

Such spectra have been investigated extensively (Segura et al. 2003, 2005; Kaltenegger et al. 2007, 2010; Macdonald 1995) specifically in the mid IR (3-24 µm) (Gardner et al. 2006). The bio-signature spectra have relatively narrow bands, whereas in the same region the parent star has the broad spectrum of a black body with broader spectral features. Therefore direct inspection of the interferogram at large enough path differences, where the transform of the spectral features resulting from the parent star have disappeared can emphasize the existence of bio-signatures. However, the only narrow spectral features expected in the parent starlight are Fraunhofer lines with λ < 1µm; therefore the starlight must be filtered so as to exclude this part of the spectrum. It is beyond the scope of this work to delve into any specific model, so as a working example we take Earth-like molecules.

The bands shown in Table 1 lead to features in the interferogram out to path differences of 1/δk ~ 3-5 mm. On the other hand the parent star, as described above, will provide no features in the interferogram at path differences greater than several wavelengths. If the combined planet-star light is filtered by a long pass filter with a gradual cut-off at around 5µm, such bio-signatures will make a major contribution to the interferogram and will become dominant at path differences greater than about 10 µm. The gradual cut-off is necessary in order to avoid introduction of high frequency features into the interferogram.

We have modeled such a system using the bio-signatures in Table 1, and also several spectral classes of parent stars. The results suggest that this method should be applicable to real planet searches. The long-pass filter is defined by an error-function with a smooth enough cut-off which was calculated not to introduce artifacts in the interferogram. In order to fulfill the Nyquist criterion, we need to sample the interferogram at intervals of half of the shortest wavelength in the bands of interest, i.e. 6.76 µm. On the other hand, the total length of path difference to be sampled in order to get maximum resolution is about 2 mm. Therefore the number of sampling channels $M$ is at most 600. However, a smaller number of channels could be used with a resultant degradation of the spectral resolution, down to a minimum of 60 which is needed to detect the longest wavelength in Table 1 with a resolving power of 10. The total photon flux is the sum of radiation from the exoplanet and parent star, emission from the telescope, atmospheric thermal radiation, and zodiacal light. The photon flux from the parent star depends on its apparent magnitude, $m$, the telescope aperture area, $s$, and the atmospheric transmittance, $T$ (unless a space telescope is employed). We wish to estimate the number of photons accumulated per channel, during an acceptable exposure time, $t$. The total number of photons $N_{phs}$ received from the star is

$$N_{phs} = 2.51^{-m} t T s I_0, \qquad (1)$$

where $I_0$ is the spectral irradiance from a zero-magnitude star in each of the standard spectral bands of the infra-red range (M-Q bands) multiplied by the corresponding wavelength range (Labeyrie et al. 2006).

The average number of photons received at any path difference in a Fourier spectrometer, $N$, assuming efficient optics and detectors in both exit ports of the interferometer, will be

$$N = N_{phs} + N_{php} + N_{tele} + N_{at} + N_{zl}, \quad (2)$$

where $N_{php}$ is the photon flux received from the planet, $N_{tele}$ is the telescope thermal flux, $N_{at}$ is atmospheric radiation, and $N_{zl}$ are photons received from zodiacal dust scattering. Combining the above equations, considering $N_{php}$ as the signal, gives a signal-to-noise in each of the $M$ channels (Barducci et al. 2011; Maillard 1987)

$$SNR = N_{php}(NM)^{-0.5}. \quad (3)$$

Now, the planet spectral signal contains both a reflection of the parent star and also its thermal self-emission with an albedo $A$, modulated by the bio-signature. For an Earth-like planet and Sun-like star (spectral class G or K) the contrast is about $10^{-6}$ for $A=0.3$ in the infra-red (M-Q bands) (Lipson et al. 2010).

Two scenarios for such a contrast were studied. The first was a ground-based telescope such as the VLT (Labeyrie et al. 2006; ESO Overview). The atmospheric noise was calculated based on the data from Lawrence et al. (2002) and the telescope emission as published for this telescope. The atmospheric transmission in each band is given by Labeyrie et al. (2006). Atmospheric emission was estimated as $m = 3$ per arcsec$^2$ in the spectral region of interest (Lawrence et al. 2002) which makes a negligible contribution if adaptive optics and a cold field stop $< 0.25$ arcsec$^2$ are used. Equation 3 shows that a $SNR$ of 2 is obtained using $M = 60$ channels, an exposure time of $t\sim10^5$s and a contrast ratio $N_{php}/N_{phs} \sim 10^{-6}$. We thus find that the limiting magnitude of the parent star for which the proposed technique would succeed is $m = 5.2$ (Lena et al. 1998; Tokunaga 2000). This gives a limiting range of about 12 parsec, or 40 light years, within which there are hundreds of stars of both classes. This estimate is based on star density statistics from our own solar neighborhood which states that about 18% of the stars within a 10 parsec range are Sun-like (Henry et al. 2011). Some of these would be favorable for sustaining conditions for an Earth-like planet.

The second scenario was a space-based telescope like the James Webb Space Telescope (JWST) (Gardner et al. 2006) for which we include the telescope emission and the Zodiacal light noise (Ootsubo et al. 2000). For this case we received a limiting parent star magnitude of $m = 5.5$ where $T=1$ and $SNR=2$. This gives a limiting range of about 14 parsec, or 46 light years.

We now turn to some simulations of the type of spectral signals expected. We chose ε Indi ($m \sim 2.2$) (Volk 2003), as an example of a class K star within the parsec range described above. The habitable zone radius for an Earth-like planet revolving around this star is 0.2 AU (Kasting et al. 1993; Kaltenegger et al. 2010). We modeled the ground-based telescope VLT with the appropriate atmospheric and telescope noise. The parameters used were $t \sim 10^4$ s, $SNR = 2$, $M = 60$ channels, atmospheric transmission $T$, modeled for each spectral band (Labeyrie et al. 2006) and $s = 212$ m$^2$.

Figure 2a shows in its entirety the interferogram obtained from the star chosen, treated as a black-body with spectral features of the Sun (Renewable Resource Data Center, Solar Spectra), after applying a long-pass error-function filter as described in the figure caption. The spectral information in the interferogram has decayed after a few periods of oscillations. We confirmed that the long pass filter makes no contribution to it. Figure 2b shows the selected region with path-difference 3-5 mm, expanded to demonstrate the photon noise from the various sources as described before. Figure 2c shows an example of the selected region in Figure 2b when a habitable planet with contrast of $10^{-6}$ is added. The planet spectrum in-cludes atmospheric-like gases and a chlorophyll molecular signature resulting from a major part of its surface.

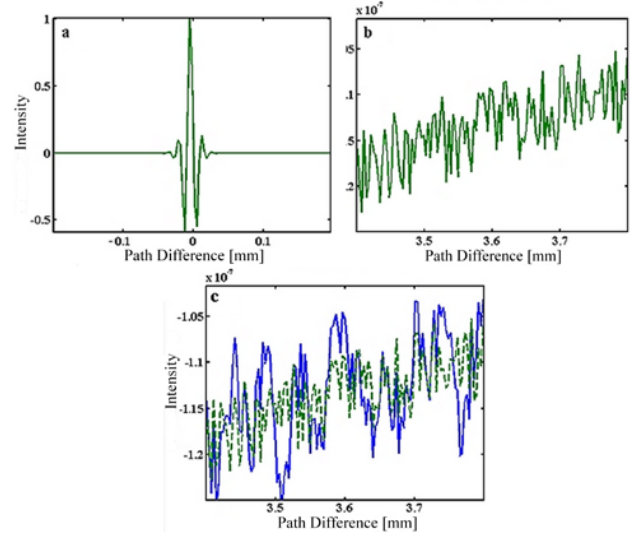

Figure 2: a) The interferogram of ε-Indi after applying a low-pass filter at 2000 cm$^{-1}$ with error function profile σ=1200 cm$^{-1}$. b) Photon noise from the various sources in the selected path difference region. Exposure time on the VLT was t$\sim 10^4$ s divided into 60 channels with SNR=2. c) Same noise as (b) including the planet bio-signature with $10^{-6}$ contrast.

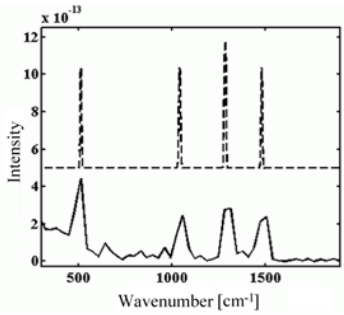

Figure 3. The bio-signature used (above) and that reconstructed from the selected region of the interferogram with photon noise (below).

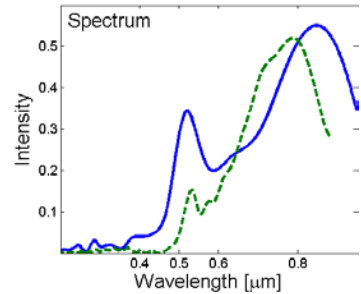

Figure 4. A fourfold contrast improvement is seen in this comparison between the spectra retrieved from the entire interferogram (green, dashed) and that retrieved from the selected part of the interferogram (blue).

To show that a bio-signature can be extracted reliably from such partial information we show in Figure 3 the power spectrum derived by Fourier analysis using 60 measurement channels from the region of the interferogram of 3-5 mm path difference. It compares well with the original bio-spectrum despite the poor signal-to-noise of the signal itself. Based on simulations of this sort, we expect this method to provide information about the spectral signatures of possible exoplanets biomolecules. Modulation of the bio-signature intensity and polarization as the planet rotates and orbits about the parent star might also be detectable and improve the SNR.

We ran an experiment using a simple breadboard Fourier spectrometer (Schwartz et al. 2011) which confirmed the ideas presented above. A 2 mm aperture was illuminated by an attenuated green LED source superimposed on black body light from a halogen lamp. A comparison between the spectra retrieved from the entire interferogram and that retrieved from the selected part of the interferogram is shown in Figure 4. Although no attempt was made to optimize the optical setup it is clear from the figure that the relative signals from the LED and the black body has been improved.

An obstacle for this method might be that, if the telescope is situated on the ground, all radiation will be modulated by the Earth's atmosphere itself. While a space telescope is the obvious solution to this, any method of significantly reducing the Earth's signature will help, such as a balloon platform or an Antarctic observatory, because of their reduced water-vapor contribution. Another possibility is to use a filter which only passes light in the known atmospheric windows where the molecular absorption by our atmosphere is negligible. Within such a window, the detection would then be restricted to the signature of chlorophyll-like and other molecules which are ground-based and must definitely be attributed to the planet.

In summary, we have proposed a method of differentiating between spectral bands of different widths by optimized sampling of their Fourier interferograms. We tested it by simulations related to detecting possible habitable extra-solar planets using their spectral bio-signatures, in the presence of overwhelming parent star radiation and different noise sources. Also, we conducted a laboratory experiment which illustrated the method using two light sources. We showed that, using adaptive optics and a cold aperture, the background radiation can be reduced sufficiently that the method will be applicable to stars with limiting magnitudes of $m = 5.2$ at a distance of 12 parsec for the VLT. The limiting magnitude in the case of the future JWST space-based telescope is $m = 5.5$ at a distance of 14 parsec. This method could also have applications outside astrophysics.

We are grateful for useful experimental assistance and inputs from Dr. D. Spector, N. Meitav and S. Hoida. E.S. acknowledges financial support from the Asher Space Institute at Technion. Parts of this work were supported by the Israel Science Foundation and the Ministry of Science.